\documentclass{ws-procs9x6}
\usepackage[latin1]{inputenc}
\usepackage[T1]{fontenc}
\usepackage{graphicx}
\usepackage[dvips]{epsfig}
\usepackage{color}

\begin{document}

\title{Quantum statistics effects for Schwinger pair production in short laser pulses} 
\author{F.~HEBENSTREIT and R.~ALKOFER}
\address{Institut f\"ur Physik, Universit\"at Graz, A-8010 Graz, Austria}
\author{G.~V.~DUNNE}
\address{Department of Physics,University of Connecticut, Storrs, CT 06269, USA}
\author{H.~GIES}
\address{Theoretisch-Physikalisches Institut, Universit\"at Jena, D-07743 Jena, Germany\\Helmholtz Institut Jena, D-07743 Jena, Germany}
\date{\today}

\begin{abstract}
We investigate non-perturbative pair production from vacuum (the Schwinger effect) in the focal region of two counter-propagating, ultra-short laser pulses with sub-cycle structure. We use the quantum kinetic formulation to calculate the momentum spectrum of created particles and show the extreme sensitivity to the laser frequency $\omega$, the pulse length $\tau$ and the carrier-envelope absolute phase $\phi$. We apply this formalism to both fermions and bosons to illustrate the influence of quantum statistics in this type of electric background field.
\end{abstract}


\bodymatter

\section{Introduction}

Non-perturbative electron-positron pair production due to the instability of the vacuum in the presence of strong external electric fields --- the so-called Schwinger effect --- has been a long-standing prediction of quantum electrodynamics (QED) \cite{sauter31,heisenberg35,schwinger51} but has not been observed yet. This effect was first considered for spatially homogeneous and static electric fields. The rate is exponentially small, with the scale set by the critical field strength which is of the order of $E_\mathrm{cr}=m^2c^3/e\hbar\approx10^{18} \, \mathrm{V}/\mathrm{m}$. While the production of constant electric fields of this order is rather unrealistic, recent developments in laser technology have raised hopes to approach the Schwinger limit in the focal region of colliding laser pulses --- either at optical high-intensity laser facilities such as ELI or in X-ray free electron laser (XFEL) systems. 

In this investigation we model the electric field produced in the focal region of two counter-propagating laser pulses by assuming that the scale of spatial variation of the electric field is much larger than the Compton wavelength. Thus, we approximate the experimental situation by a spatially homogeneous electric field $\vec{E}(t)=(0,0,E(t))$, represented by an oscillatory field with a temporal Gaussian envelope:
\begin{equation}
  \label{eqn:elfield}
  E(t)=E_0\cos(\omega t+\phi)\exp\left(-\frac{t^2}{2\tau^2}\right)\ .
\end{equation}
For this type of electric field there is a simple analytic expression of the time-dependent vector potential $\vec{A}(t)=(0,0,A(t))$ in terms of complex error functions \cite{hebenstreit09}. Due to the appearance of such a variety of physical parameters -- the field strength $E_0$, the laser freqency $\omega$, the pulse length parameter $\tau$ and the carrier-envelope absolute phase (carrier phase) $\phi$ -- we are faced with a rather complicated interplay between various scales, which ultimately leads to distinctive signatures in the momentum distribution of produced pairs \cite{hebenstreit08,hebenstreit09}. 

\section{Quantum Kinetic Equation}

The Schwinger effect is a non-equilibrium, time-dependent quantum process and hence quantum kinetic theory provides an appropriate framework. The quantum kinetic formulation arises as a rigorous connection between kinetic theory and mean-field approximation to scalar QED (sQED) and QED \cite{kluger98,schmidt98}. The key quantity in this approach is the momentum distribution function $f_\pm(\vec{k},t)$ which satisfies a non-Markovian quantum Vlasov equation including a source term for particle-antiparticle pair production:
\begin{equation}
  \frac{\mathrm{d}}{\mathrm{d}t}f_\pm(\vec{k},t)=\frac{W_\pm(t)}{2}\int\limits_{-\infty}^{t}{\mathrm{d}t'W_\pm(t') \left[1\pm2f_\pm(\vec{k},t')\right]\cos\left[2\int_{t'}^{t}{\mathrm{d}t''\,\omega(t'')}\right]} \ .
\end{equation}
Denoting bosons with $(+)$ and fermions with $(-)$, $W_\pm(t)$ are given by
\begin{equation}
  W_+(t)=\frac{eE(t)p_\parallel(t)}{\omega^2(t)} \qquad \mathrm{and} \qquad W_-(t)=\frac{eE(t)\epsilon_\perp}{\omega^2(t)} \ , 
\end{equation}
with $e$ being the electric charge. $\vec{k}=(\vec{k}_\perp,k_\parallel)$ is the canonical three-momentum vector and  $p_\parallel(t)=k_\parallel-eA(t)$ is the kinetic momentum along the electric field direction. $\epsilon_\perp^2 = m^2+\vec{k}_\perp^{\,2}$ is the transverse energy squared and $\omega^2(t)= \epsilon_\perp^2+p_\parallel^2(t)$ characterizes the total energy squared. It is absolutely crucial to note that $f_\pm(\vec{k},t)$ has physical meaning as the distribution function of real particles only at asymptotic times $t\rightarrow \pm\infty$.

\section{Quantum Statistics Effect}

\begin{figure}[th]
  \begin{minipage}[t]{5.4 cm}
    \centering
    \includegraphics[width=6cm,height=3.5cm]{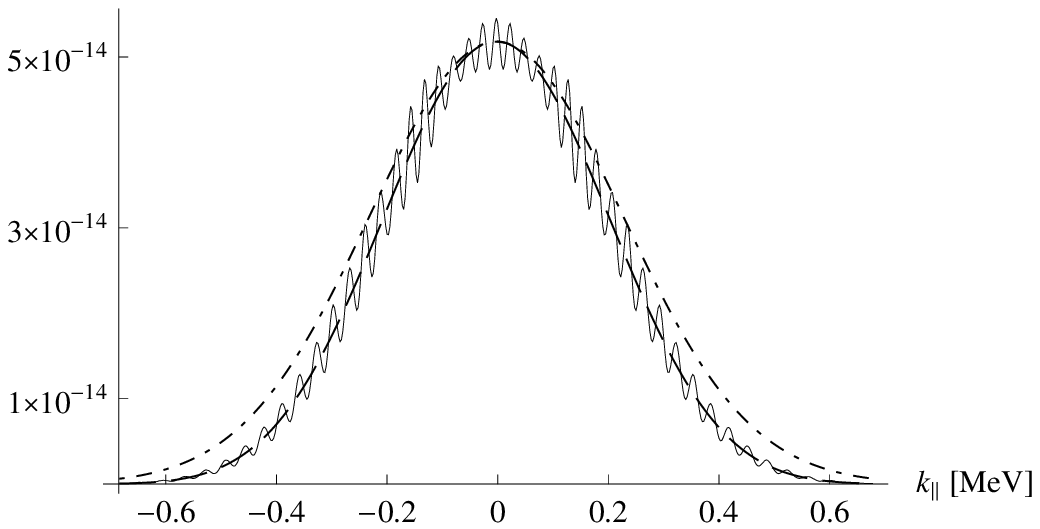}
  \end{minipage}
  \begin{minipage}[t]{5.4 cm}
    \centering
    \includegraphics[width=6cm,height=3.5cm]{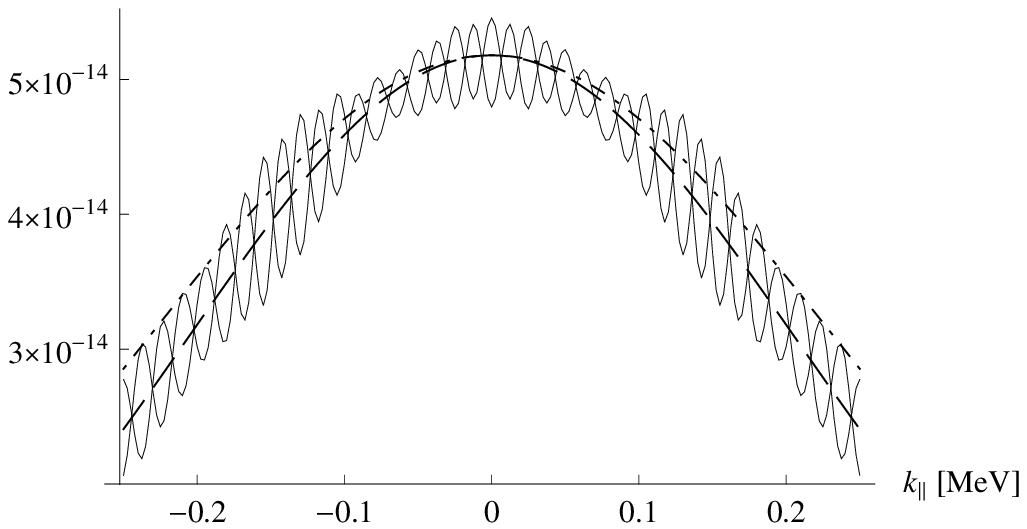} 
  \end{minipage}
  \caption{\label{fig:statistics1} Carrier phase $\phi=0$; Left: Momentum distribution function $f_-(\vec{k},\infty)$ for $\vec{k}_\perp=0$ (solid line) in comparison with the WKB-Gaussian approximation Eq.~(\ref{eqn:popovmomdist}) (dashed-dotted line) and the WKB instanton result Eq.~(\ref{eqn:kimmomdist}) (dashed line). Right: Momentum distribution functions $f_\pm(\vec{k},\infty)$ for $\vec{k}_\perp=0$ in more detail (solid lines). At momentum values, where QED predicts a local maximum, sQED predicts a local minimum and vice versa. Note that the WKB instanton result Eq.~(\ref{eqn:kimmomdist}) (dashed line) lies almost perfectly on the intersection points, while the WKB-Gaussian approximation (\ref{eqn:popovmomdist}) does not.}
\end{figure}

We consider the subcritical field strength regime $E_0=0.1E_\mathrm{cr}$, with $\tau=2\cdot10^{-4}\, \mathrm{eV^{-1}}$, first concentrating on the case of vanishing carrier phase $\phi=0$. It has been shown in a previous publication \cite{hebenstreit09} that the momentum distribution function $f_-(\vec{k},t)$ in QED exhibits a distinctive oscillatory structure for $\sigma\equiv\omega\tau\gtrsim4$, with the oscillation scale set by the laser frequency $\omega$. An analogous calculation in the framework of sQED gives a very similar result. However, due to the difference in quantum statistics, $f_-(\vec{k},\infty)$ shows a local maximum at momentum values at which  $f_+(\vec{k},\infty)$ shows a local minimum, and vice versa, as shown in Fig.~\ref{fig:statistics1}.

The Schwinger effect in the electric field (\ref{eqn:elfield}) without carrier-phase $\phi$ has been investigated previously in the framework of a WKB approximation, together with a Gaussian approximation for the momentum distribution \cite{popov01}:
\begin{equation}
  \label{eqn:popovmomdist}
  \frac{\mathrm{d}^3\mathcal{P}}{dk^3}\sim\exp\left(-\frac{1}{eE_0}\left[\frac{1+\sigma^2}{\sigma^2}\gamma^2\,k_\parallel^2+\vec{k}_\perp^2\right]\right) \ ,
\end{equation}
with $\gamma\equiv m\omega/eE_0$ being the Keldysh parameter. In fact, this approximation is too crude in several aspects: First, it does not see the distinctive oscillatory structure found in the exact (numerical) treatment; second, the Gaussian shape is somewhat broader than the true distribution function. In order to explain this discrepancy, we apply the quantum mechanical WKB instanton method \cite{kim07}, for which the momentum distribution is 
\begin{equation}
  \label{eqn:kimmomdist}
  \frac{\mathrm{d}^3\mathcal{P}}{dk^3}\sim\exp\left(-2S_{\vec{k}}\right) \ \ \mathrm{with} \ \ 2S_{\vec{k}}=i\oint_{\Gamma}\sqrt{m^2+\vec{k}_\perp^2+[k_\parallel-eA(t)]^2}\,\mathrm{d}t \ ,
\end{equation}
with $\Gamma$ being the contour around the branch cut. After a change of variable, from $t$ to $T=-A(t)/E_0$ we expand the instanton action and obtain an approximate solution in terms of an infinite series in powers of the dimensionless variables $\epsilon=\epsilon_\perp/(eE_0\tau)$ and $\kappa=k_\parallel/(eE_0\tau)$: 
\begin{equation}
  2S_{\vec{k}}=\frac{\pi\epsilon_\perp^2}{eE_0}\sum_{i=0}^{\infty}{S^{(2i)}} \ ,
\end{equation}
with the first term $S^{(0)}=1$, and the next three terms being given by
\begin{eqnarray}
   S^{(2)}&=&\left[\kappa^2-\frac{\epsilon^2}{4}\right]\frac{1+\sigma^2}{2} \ ,\\ 
   S^{(4)}&=&\left[\kappa^4-\frac{3\kappa^2\epsilon^2}{2}+\frac{\epsilon^4}{8}\right]\frac{7+14\sigma^2+9\sigma^4}{24} \ , \\
   S^{(6)}&=&\left[\kappa^6-\frac{15\kappa^4\epsilon^2}{4}+\frac{15\kappa^2\epsilon^4}{8}-\frac{5\epsilon^6}{64}\right]\frac{127+381\sigma^2+463\sigma^4+225\sigma^6}{720}
\end{eqnarray}
It is straightforward to calculate even higher order terms. But taking the first four terms into account, Eq.~(\ref{eqn:kimmomdist}) already agrees very well with the averaged envelope of the exact momentum distribution. We point out that this averaged envelope lies almost perfectly on the intersection points between the momentum distribution of scalar and spinor particles, as shown in Fig.~\ref{fig:statistics1}. Note that none of the semiclassical estimates predicts the oscillatory structure in the momentum distribution.

\vspace{0.2cm}

\begin{figure}[ht]
    \begin{minipage}[t]{5.4 cm}
    \centering
    \includegraphics[width=6cm,height=3.5cm]{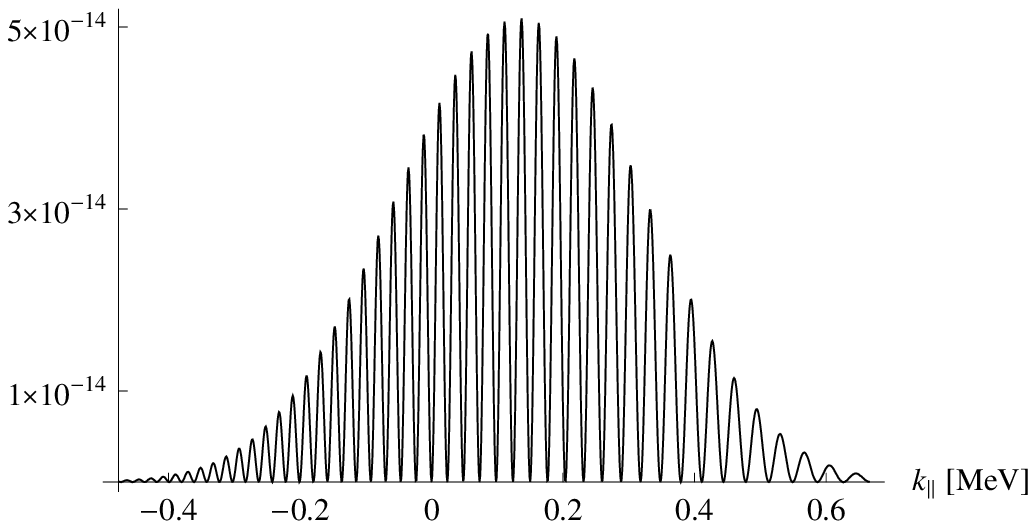}
    \end{minipage}
    \begin{minipage}[t]{5.4 cm}
    \centering
    \includegraphics[width=6cm,height=3.5cm]{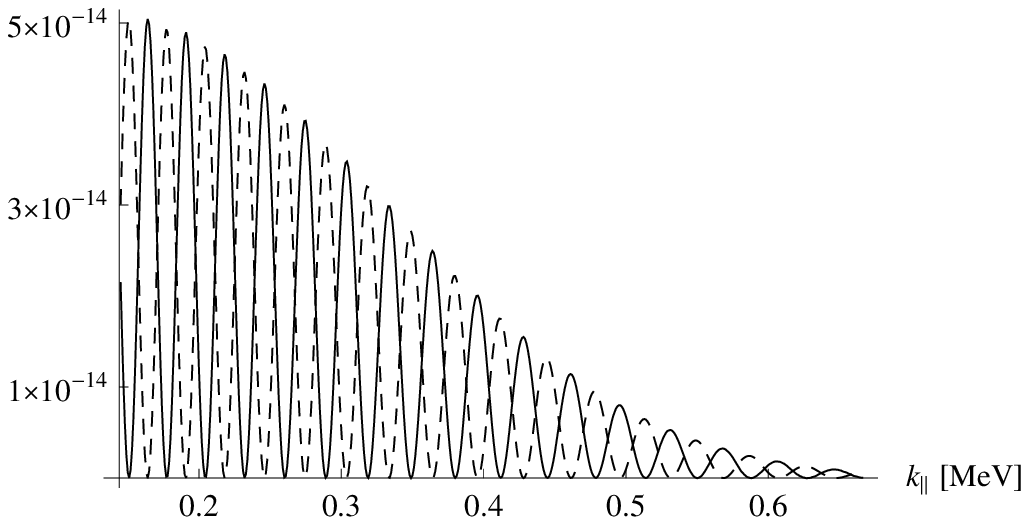}
    \end{minipage}
  \caption{\label{fig:statistics2} Carrier phase $\phi=-\pi/2$; Left: Momentum distribution function $f_-(\vec{k},\infty)$ for $\vec{k}_\perp=0$. Right: Comparison of the momentum distribution functions $f_-(\vec{k},\infty)$ (solid line) and $f_+(\vec{k},\infty)$ (dashed line), in more detail.}
\end{figure}

Finally, considering a carrier phase shift $\phi=-\pi/2$, there are momentum values at which no particles are expected to be produced \cite{hebenstreit09}, which is due to a resonance phenomenon in the equivalent scattering picture \cite{dumlu09}. We emphasize that this behavior can be expected for any time-antisymmetric electric field, corresponding to a time-symmetric vector potential $A(t)$. Again, comparing the results for scalar and spinor particles, we observe that $f_-(\vec{k},\infty)$ shows a local maximum at momentum values at which $f_+(\vec{k},\infty)$ shows a local minimum, and vice versa, as shown in Fig.~\ref{fig:statistics2}. 

\section{Summary}

The momentum distribution of produced particles is extremely sensitive to the physical parameters of a laser pulse. The same qualitative behavior is obtained for both scalar and spinor particles, but, due to quantum statistics, the oscillatory structure is interchanged. This new effect is complementary to the interchange of statistics found in the analogue-thermal distribution properties of the QED effective action in electric field backgrounds \cite{muller77,pauchyhwang09}.

Here we have considered two identical, colliding laser pulses. Recent proposals consider more complicated situations to overcome the strong suppression of the Schwinger effect \cite{schutzhold08,piazza09,dunne09}. Applying the quantum kinetic formulation, it would be possible to determine not only the total rate but also the momentum distribution.

\vspace{0.4cm}

\noindent {\bf Acknowledgments:} We acknowledge support from the DOC program of the Austrian Academy of Sciences and from the FWF doctoral program DK-W1203 (FH), from the US DOE grant DE-FG02-92ER40716 (GD), and from the DFG grant Gi328/5-1 and SFB-TR18 (HG).


\begin{thebibliography}{9}

\bibitem{sauter31} F.~Sauter, {\em Z.~Phys.} {\bf 69}, 742 (1931). 

\bibitem{heisenberg35} W.~Heisenberg and H.~Euler, {\em Z.~Phys.} {\bf 98}, 714 (1935).

\bibitem{schwinger51} J.~Schwinger, {\em Phys.~Rev.} {\bf 82}, 664 (1951).

\bibitem{hebenstreit08} F.~Hebenstreit, R.~Alkofer and H.~Gies, {\em Phys.~Rev. D} {\bf 78}, 061701 (2008).

\bibitem{hebenstreit09} F.~Hebenstreit et al., {\em Phys.~Rev.~Lett.} {\bf 102}, 150404 (2009).

\bibitem{kluger98} Y.~Kluger, E.~Mottola and J.~M.~Eisenberg, {\em Phys.~Rev. D} {\bf 58}, 125015 (1998).

\bibitem{schmidt98} S.~Schmidt et al., {\em Int.~J.~Mod.~Phys. E} {\bf 7}, 709 (1998).

\bibitem{popov01} V.~S.~Popov, {\em JETP Lett.} {\bf 74}, 133 (2001).

\bibitem{kim07} S.~P.~Kim and D.~N.~Page, {\em Phys.~Rev. D} {\bf 75}, 045013 (2007).

\bibitem{dumlu09} C.~Dumlu, {\em Phys.~Rev. D} {\bf 79}, 065027 (2009).

\bibitem{muller77} B.~Muller, W.~Greiner and J.~Rafelski, {\em Phys.~Lett. A} {\bf 63}, 181 (1977).

\bibitem{pauchyhwang09} W.~Y.~Pauchy Hwang and S.~P.~Kim, {\em Phys.~Rev. D} {\bf 80}, 065004 (2009).

\bibitem{schutzhold08} R.~Sch\"utzhold, H.~Gies and G.~V.~Dunne, {\em Phys.~Rev.~Lett.} {\bf 101}, 130404 (2008).

\bibitem{piazza09} A.~Di Piazza et al., {\it arXiv:0906.0726} (2009).

\bibitem{dunne09} G.~V.~Dunne, H.~Gies and R.~Sch\"utzhold, {\it arXiv:0908.0948} (2009).


\end{thebibliography}
\end{document}